# A Fully Reconfigurable All-Optical Integrated Nonlinear Activator

**Authors:**

Bei Chen[1], Xiaowen Xiong[2], Renheng Zhang[1], Yitang Dai[3], Jianyi Yang[4], Jinhua Bai[1], Wei Li[1], Ninghua Zhu[1,5*], Ming Li[1,6*]

**Affiliations:**

[1] Key Laboratory of Optoelectronic Materials and Devices, Institute of Semiconductors, Chinese Academy of Sciences, Beijing 100083, China

[2] Department of Electronic Engineering, Tsinghua University, Beijing 100084, China

[3] State Key Laboratory of Information Photonics and Optical Communications, Beijing University of Posts and Telecommunications, Beijing 100876, China

[4] College of Information Science and Electronic Engineering, Zhejiang University, Hangzhou 310027, China

[5] Institute of Intelligent Photonics, Nankai University, Tianjin 30071, China

[6] College of Materials Science and Opto-Electronic Technology, University of Chinese Academy of Sciences, Beijing 100049, China

**E-mail:**

*e-mail: nhzhu@semi.ac.cn, ml@semi.ac.cn

**Abstract:**

**Photonic neural networks have been considered as the promising candidates for next-generation neuromorphic computation, aiming to break both the power consumption wall and processing speed boundary of state-to-date digital computing architectures. Optics has shown its advantages in parallelism and linear manipulation. However, the lack of low-power and high-speed all-optical nonlinear activation neurons limits its revolution in large-scale photonic neural networks. Here we demonstrate an all-optical nonlinear activator (AONA) based on Fano-enhanced nonlinear optical effects in intra-cavity field, in which our device enables reconfigurability both in shape and type of the nonlinear functions (NFs) relying on the tuning biases on Fano interference and cavity buildup. Experimental results show that our AONA enables nonlinear optical computing with low-power**

**continuous-wave light of 0.1 mW threshold, which can also support the overall processing speed of 13 GHz. The performances of the generated reconfigurable NFs are verified in the classification of handwritten digits and image recognition tasks, yielding the converged cost and enhanced accuracy compared to the linear-only networks. Our proposed device would pave the way for energy-efficient and accelerated all-optical intelligent processors with versatile functionalities and large-scale integration.**

Neuromorphic photonics has emerged as a promising solution[1, 2] to overcome the great constraints of traditional electronic hardware for the escalating computing consumption in the coming exaflop era, primarily due to the electronic natural bottleneck of processing speed, power consumption and parallel capability. Relying on the superiorities of silicon photonics[3, 4], various silicon-based integrated photonic neural network architectures[5-14] have been proposed to execute high-speed linear transformations with low-power consumption, especially for matrix-vector multiplications (MVMs) or convolutional computing. Such linear functions have been realized by cascaded Mach-Zehnder interferometer (MZI) arrays[5-7], microring resonator (MRR) weighted banks[8-10], multimode interference (MMI) couplers[11, 12], diffractive metalines[13, 14], etc. However, silicon-based integrated photonic neural networks still face challenges in large-scale integration and versatile on-chip functionalities due to the lack of all-optical nonlinear activators (AONAs). The generated nonlinear functions (NFs), similar to the function of the neural synapses in the real brain[15], are indispensable for dealing with complicated tasks. They are essential to accelerate the convergence of the network and improve the computational accuracy[16, 17]. Moreover, in order to face versatile practical scenarios, numerous types of NFs have been searched to replace the widely used rectified linear unit (ReLU)[18] or Sigmoid function, and are required to be adaptively adopted in different models and datasets[19, 20].

The reported nonlinear computations in photonic neural networks are mainly based on optoelectronic activation structures[21-25] or electronic computation[12], resulting in drawbacks of high-power consumption, limited processing speed, analog-to-digital conversion latency and non-programmable optoelectronic nonlinearity. Considering complementary metal-oxide-semiconductor (CMOS) compatibility between linear and nonlinear operations, a significant challenge of AONAs is the intrinsically weak nonlinear effect of silicon material[26]. In order to enhance the nonlinear effect, one solution is to monolithically transfer particular optical nonlinear materials on silicon, such as phase-change materials (PCMs)[10, 27] and two-dimensional materials[28, 29], which directly impact on the refractive index or absorption coefficient of the devices by the optical power of input signals. However, their stabilities cannot be guaranteed due to the specific manufacture process[30] or material characteristics[31, 32], hindering the large-scale integration with silicon photonics. Meanwhile, these nonlinear responses are generated with relatively high optical power thresholds, resulting in large power consumption. In addition, the nonlinear processing speed is constrained by the response time of the changes in material properties, such as the crystallization of PCMs. The other solution is to directly fabricate nonlinear activators on standard silicon photonic platforms, relying on germanium-silicon (Ge/Si) structures[33-35] or cavity-enhanced nonlinear effects[36, 37]. The nonlinear Ge/Si photodiodes[33], in which a strong all-optical nonlinear response is produced by the accumulated carriers and the free carrier absorption (FCA) in Ge film, can realize much more high processing speed of GHz level, compared with Ge/Si hybrid passive waveguides[34, 35] with that up to MHz level. However, without tunable transmission spectrum generated by cavity structures[36, 37], these nonlinear responses are unchangeable, lacking flexibility and reconfigurability. Therefore, there is an urgent requirement to find a flexible solution to implement AONAs with enriched functional versatilities, lowered nonlinear threshold and enhanced processing speed, which would harness low-power, high-speed and scalable all-optical computing with continuous-wave input light.

In this work, we propose and demonstrate a fully CMOS-compatible AONA based on an MZI-embedded MRR structure with Fano interference. The intra-cavity buildup enhanced by Fano resonances allows efficient nonlinear operations. Paired with the tuning biases on Fano interference and the embedded MZI in cavity, the generated NFs are flexible and reconfigurable in both the curve shape and the function type. Meanwhile, both the static characterization and the high-speed responses show that a total of six classes of different NFs can be realized. Experimentally, nonlinear optical computing can be triggered by the continuous-wave inputs with the optical power of 0.1 mW. In addition, the nonlinear power-in and power-out relationship can be captured, as the continuous-wave inputs are modulated with 13 GHz radio-frequency signals. To illustrate the feasibility of the generated NFs, two image classification tasks using MNIST and CIFAR-10 datasets are challenged. The results demonstrate that each class of the generated NFs enables faster convergence speed and better accuracy, compared to the linear-only networks. Overall, our findings showcase the enhanced ability of the proposed Fano-enhanced AONA structure for large-scale all-optical computing with low power consumption, high processing speed and versatile functionalities.

## Device design and feature characterization

Our proposed AONA device (Fig. 1a) is fabricated on an 8-inch silicon-on-insulator wafer with 220-nm-thick silicon layer and 2-μm-thick buried oxide layer by Advanced Micro Foundry (AMF). Detailed device geometry and optical field information can be found in Section II in Supplementary Information. To measure the nonlinear threshold and the processing speed of the proposed AONA device, two experimental setups (Fig. 1a) are carried out, including (1) static characterization by measuring the optical power of device's output ports versus variable optical input power of continuous-wave laser (Exp. I and Supplementary Fig. S4), and (2) instantaneous NF generation under high-speed responses by the modulation of sinusoidal input signal with different frequencies (Exp. II and Supplementary Fig. S6).

The $P_{out}$-$P_{in}$ relations under static states are shown in Fig. 1b, in which only input wavelength and the bias on Fano interference (using a two-dimensional parameter sweep) are tunable. Meanwhile, the intra-cavity buildup is kept with the best quality under the bias on embedded MZI of $\boldsymbol{\pi}$. Considering evitable on-chip thermal crosstalk, we convert the biases on MRR to the tunable input wavelengths and adopt a thermo-electric cooler module to maintain the on-chip temperature. The measured $P_{out}$-$P_{in}$ relations can be sorted into six types of different NFs (including Softplus, ReLU, Fourier, Sine, Rational and Clipped ReLU) and further fitted with the corresponding mathematical expressions (Supplementary Table S2 and Fig. S5), which can all be related to the commonly used activation functions or radial bias functions[38]. The solid dots are the average results from five repeated samplings. The thresholds of all the generated NFs are around 0.1 mW. Such a low nonlinear threshold means low power consumption, which would be significant in large-scale integration and robust for the co-design of linear and nonlinear optical computations. Besides, it is crucial to point out that within a function type, the shape and nonlinear threshold can be tuned by varying the two-dimensional parameter sweep. In addition, instantaneous all-optical NFs are captured by the mapping of input and output time-domain waveforms monitored on the sampling oscilloscope. In Fig. 1c, we plot the measured output time-domain waveforms of the first three periods of each modulation frequency (100 MHz~10 GHz). The mapped $P_{out}$-$P_{in}$ relations (Fig. 1d) representing instantaneous NFs, are also fitted (Supplementary Table S3 and Fig. S6) and consistent with the generated NFs under static characterization. The dynamic experimental results show that our proposed AONA provides a maximum processing speed of 13 GHz, in which all classes of NFs can still be generated.

Subsequently, to further comprehend and characterize the reconfigurability of NFs generated in our AONA device, the tuning process of $P_{out}$-$P_{in}$ relations under the two-dimensional parameter sweep of input wavelength and the bias on Fano interference are measured, together with a modulation frequency of GHz level (2~13 GHz, step=1 GHz).

As the heater voltage bias on Fano interference is increased from 1.5 V to 7.5 V, transmission spectra at the device's two output ports are measured and depicted in Fig. 2a. At the end voltages, extinction ratios (ERs) of transmission spectra become largest. The top row in Fig. 2b shows that instantaneous $P_{out}$-$P_{in}$ relations at output port 1 are reconfigured with the tunable input wavelengths, in which the bias on Fano interference is fixed at 1.5 V and the modulation frequency is set as 10 GHz. The tunable relations are changed from linear region to nonlinear region and came back to linear region, which indicates that under each two-dimensional parameter, operating nonlinear spans (the supporting input wavelengths) are different. Besides, the bottom row shows the corresponding changing process of $P_{out}$-$P_{in}$ functions at output port 2, in which the NFs' classes are generated with an opposite sequence. Meanwhile, these two output ports provide partly overlapping operating nonlinear spans. Such results mean that this AONA device can generate two different NFs or one NF and one linear function or two linear functions at the same time, providing the functional versatility for different neural network models and the possibility of max-pooling dropout operation[39] for networks' regulation. In addition, we evaluate operating nonlinear spans for NFs' generation of two output ports under different Fano biases. Figure 2c shows that at the end voltages on Fano interference, nonlinear spans are all increased from 2 GHz to 10 GHz and then gradually decreased. Besides, as shown in Fig. 2d, the maximum nonlinear spans at each applied Fano bias (measured with each modulation frequency from 2~13 GHz) are with a positive correlation of the corresponding ERs. Consequently, a full look-up table of operating nonlinear spans at two output ports is illustrated in Fig. 2e, which demonstrates the complete characteristics of AONA device. It is noticed that our AONA device will not provide nonlinear region when the applied Fano bias is set between 5 V to 5.5 V, due to the weak nonlinearity lying in the intra-cavity buildup.

### Demonstration with handwritten digit and CIFAR-10 image classification

To validate the performances of the generated NFs in our AONA device, two classification tasks including handwritten digits and image recognition based on the MNIST and CIFAR-10 dataset are adopted. Correspondingly, two neural network models based on three-layer fully connected neural network (FCNN) with back-propagation algorithms[40] and LeNet-5[41] are depicted in Figs. 3a and 3c, respectively. The MNIST database with 28x28 gray pixels has a training set of 60,000 examples and a test set of 10,000 examples. In addition, RGB-colored images of 32x32 pixels in the CIFAR-10 database contain 50,000 training images together with 10,000 test images.

In both training models, all the nonlinear activation functions are replaced by our generated optical NFs and the cross-entropy loss is adopted as the loss function. Meanwhile, these two models are also trained without any nonlinear activation functions (linear-only) as the baseline for comparison. The convergence cost and accuracy of two classification tasks with different optical NFs are demonstrated in Figs. 3b and 3d. In the MNIST database, our optical NFs all show a much faster convergence speed and a lower cost, indicating their better convergence ability than the linear-only model. Besides, six classes of our optical NFs have a cost advantage of at least 0.105 and a more than 3.59% accuracy increment. Moreover, these optical NFs also show lower convergence costs and higher accuracies in the CIFAR-10 database. The confusion matrices of each optical NFs in two classification tasks are illustrated in Section IV in Supplementary Information. These results verify that all the generated NFs produced by our proposed AONA device have higher performances on representative classification tasks.

## Discussion

Till now, the full potential of neuromorphic photonics hasn't been exploited, one key challenge being a lack of effective nonlinear operations. The essential characteristics of the effective optical NFs include: (1) Continuous wave inputs. Although optical NFs triggered by continuous wave[33-37] or pulsed[27, 42, 43] inputs are both reported, continuous modes are more suitable for various models rather than pulsed inputs mainly for spiking neural networks[44]. (2) All-optical realization. Photoelectric conversion and analog-to-digital conversion can be avoided. Hence, no extra power consumption or time latency will be brought in. (3) Low nonlinear threshold. The direct benefit is the limited power consumption for large-scale integration. Another impact is that it allows more space for insertion losses caused by linear operations and can effectively reduce the times of on-chip amplification. (4) High processing speed. A higher processing speed directly determines a faster computing speed for overall optical processors, where electric neural networks are usually restricted to the clock rate of GHz level[45]. (5) Reconfigurability. If the curve shape and the function type are both tunable, the generated optical NFs can be perfectly chosen and adapted according to the training results. Therefore, more models and datasets can be realized by using the same fabricated structures. Other features, such as CMOS compatibility, are also considered for the co-design of optical linear and nonlinear operations.

Our study presents a Fano-enhanced cavity for NFs' generation, which effectively satisfies all the above-mentioned characteristics and possesses enhanced performances. To illustrate this, the experimentally demonstrated performances of silicon-based AONA structures with continuous wave inputs are listed in Fig. 4. Meanwhile, more detailed comparisons on various reported on-chip optical nonlinear activators are given in Supplementary Table S4. One of the key advantages of our proposed AONA is the reconfigurability in both the curve shape and the function type. These generated optical NFs include various types of nonlinear activation function and radial basis function. It indicates that the optical NFs could be adjusted and adaptive for the requirement of different practical scenarios, which greatly differs from the previous works[33, 34] with only one unchangeable NF. Experimental results show the complete generation process for six classes of different NFs, which all bring in improved accuracies of at least 3.59% and 3.88% for handwritten recognition and image classification, respectively. Moreover, the generated functions at two output ports of this AONA device can be different, such as two linear functions, a linear and another NF, two NFs in the same function type or different types. Such results enrich the device's functional versatility. In future work, we could further use it in large-scale complicated networks as the flexible optical nonlinear activators together with the function of max-pooling dropout. Another obvious advantage of the AONA is simultaneously possessing a high processing speed and a low nonlinear threshold, satisfying the development trend of optical nonlinear operations. Such device can support an overall computing speed of $1.3\times10^{10}\times L$ operations/s where L is determined by the architectures of optical linear operations.

Overall, the Fano-enhanced cavity presents a novel approach for generating on-chip all-optical nonlinearity with high processing speed under low-power continuous wave inputs, yielding reconfigurability in both curve shapes and function types. We also demonstrate the generation process of the optical NFs together with tunable nonlinear spans within different intra-cavity buildup, in which the degree of on-chip nonlinearity is adjusted with the biases of Fano interference. These NFs are all trained for FCNN and LeNet-5 models with enhanced performances compared to the linear-only networks. Note that this AONA device is applicable to the cascade of coherent optical linear systems. These characteristics make it a promising approach for realizing integrated all-optical computing.

**Acknowledgements**

This work was supported by the National Key Research and Development Program of China under grant nos. 2023YFB2806503 (B.C.) and 2023YFB2806000 (B.C.), the Beijing Municipal Natural Science Foundation under grant no. Z210005 (M.L.), and the China Postdoctoral Science Foundation under grant nos. BX20230172 (X.X.) and 2023M731884 (X.X.).

## Author contributions

B.C. conceived and initiated the project. B.C. designed the devices and conducted the experiments. R. Z. assisted with the test. B.C. and X.X developed the training models and relevant processing code. B.C., Y. D. and J.Y. analyzed and deduced the theory. J.B. dealt with the package and thermal management of the chip. W.L. assisted with the package. All authors contributed to the discussion of experimental results and reviewed the manuscript. B.C. wrote the paper with contributions from all co-authors. N.Z. and M.L. supervised the project.

## Conflict of interest

The authors declare no conflicts of interest.

## Data availability

Data underlying the results presented in this paper are not publicly available at this time but may be obtained from the authors upon reasonable request.

## References

1. Wetzstein G, Ozcan A, Gigan S, Fan S, Englund D, Soljacic M*, et al.* Inference in artificial intelligence with deep optics and photonics. *Nature* 2020, **588**(7836)**:** 39-47.
2. Shastri BJ, Tait AN, Ferreira de Lima T, Pernice WHP, Bhaskaran H, Wright CD*, et al.* Photonics for artificial intelligence and neuromorphic computing. *Nature Photonics* 2021, **15**(2)**:** 102-114.
3. Siew SY, Li B, Gao F, Zheng HY, Zhang W, Guo P*, et al.* Review of Silicon Photonics Technology and Platform Development. *Journal of Lightwave Technology* 2021, **39**(13)**:** 4374-4389.
4. Shekhar S, Bogaerts W, Chrostowski L, Bowers JE, Hochberg M, Soref R*, et al.* Roadmapping the next generation of silicon photonics. *Nat Commun* 2024, **15**(1)**:** 751.
5. Shen Y, Harris NC, Skirlo S, Prabhu M, Baehr-Jones T, Hochberg M*, et al.* Deep learning with coherent nanophotonic circuits. *Nature Photonics* 2017, **11**(7)**:** 441-446.
6. Zhang H, Gu M, Jiang XD, Thompson J, Cai H, Paesani S*, et al.* An optical neural chip for implementing complex-valued neural network. *Nat Commun* 2021, **12**(1)**:** 457.
7. Zhu HH, Zou J, Zhang H, Shi YZ, Luo SB, Wang N*, et al.* Space-efficient optical computing with an integrated chip diffractive neural network. *Nat Commun* 2022, **13**(1)**:** 1044.
8. Tait AN, Wu AX, de Lima TF, Zhou E, Shastri BJ, Nahmias MA*, et al.* Microring Weight Banks. *IEEE Journal of Selected Topics in Quantum Electronics* 2016, **22**(6)**:** 312-325.
9. Tait AN, de Lima TF, Zhou E, Wu AX, Nahmias MA, Shastri BJ*, et al.* Neuromorphic photonic networks using silicon photonic weight banks. *Sci Rep* 2017, **7**(1)**:** 7430.

10. Feldmann J, Youngblood N, Wright CD, Bhaskaran H, Pernice WHP. All-optical spiking neurosynaptic networks with self-learning capabilities. *Nature* 2019, **569**(7755)**:** 208-214.
11. Qu Y, Zhu H, Shen Y, Zhang J, Tao C, Ghosh P*, et al.* Inverse design of an integrated-nanophotonics optical neural network. *Sci Bull (Beijing)* 2020, **65**(14)**:** 1177-1183.
12. Meng X, Zhang G, Shi N, Li G, Azana J, Capmany J*, et al.* Compact optical convolution processing unit based on multimode interference. *Nat Commun* 2023, **14**(1)**:** 3000.
13. Wang Z, Chang L, Wang F, Li T, Gu T. Integrated photonic metasystem for image classifications at telecommunication wavelength. *Nature Communications* 2022, **13**(1).
14. Fu T, Zang Y, Huang Y, Du Z, Huang H, Hu C*, et al.* Photonic machine learning with on-chip diffractive optics. *Nat Commun* 2023, **14**(1)**:** 70.
15. Rosenblatt F. *The perceptron, a perceiving and recognizing automaton Project Para*. Cornell Aeronautical Laboratory, 1957.
16. Zhengbo C, Lei T, Zuoning C. Research and design of activation function hardware implementation methods. *Journal of Physics: Conference Series* 2020, **1684**(1)**:** 012111.
17. Destras O, Le Beux S, De Magalhães FG, Nicolescu G. Survey on Activation Functions for Optical Neural Networks. *ACM Computing Surveys* 2023, **56**(2)**:** 1-30.
18. Krizhevsky A, Sutskever I, Hinton GE. Imagenet classification with deep convolutional neural networks. *Advances in neural information processing systems* 2012, **25**.
19. Prajit Ramachandran∗ BZ, Quoc V. Le. SEARCHING FOR ACTIVATION FUNCTIONS. *arXiv:171005941v2* 2017.
20. Dubey SR, Singh SK, Chaudhuri BB. Activation functions in deep learning: A comprehensive survey and benchmark. *Neurocomputing* 2022, **503:** 92-108.
21. Williamson IAD, Hughes TW, Minkov M, Bartlett B, Pai S, Fan S. Reprogrammable Electro-Optic Nonlinear Activation Functions for Optical Neural Networks. *IEEE Journal of Selected Topics in Quantum Electronics* 2020, **26**(1)**:** 1-12.
22. Tait AN, Ferreira de Lima T, Nahmias MA, Miller HB, Peng H-T, Shastri BJ*, et al.* Silicon Photonic Modulator Neuron. *Physical Review Applied* 2019, **11**(6).
23. Pappas C, Kovaios S, Moralis-Pegios M, Tsakyridis A, Giamougiannis G, Kirtas M*, et al.* Programmable Tanh-, ELU-, Sigmoid-, and Sin-Based Nonlinear Activation Functions for Neuromorphic Photonics. *IEEE Journal of Selected Topics in Quantum Electronics* 2023, **29**(6: Photonic Signal Processing)**:** 1-10.
24. Zhong C, Liao K, Dai T, Wei M, Ma H, Wu J*, et al.* Graphene/silicon heterojunction for reconfigurable phase-relevant activation function in coherent optical neural networks. *Nat Commun* 2023, **14**(1)**:** 6939.
25. Amin R, George JK, Sun S, Ferreira de Lima T, Tait AN, Khurgin JB*, et al.* ITO-based electro-absorption modulator for photonic neural activation function. *APL Materials* 2019, **7**(8)**:** 081112.
26. Cheng C-H, Fu C-S, Wang H-Y, Set SY, Yamashita S, Lin G-R. Review on optical nonlinearity of group-IV semiconducting materials for all-optical processing. *APL*

*Photonics* 2022, **7**(8): 081101.

27. On-chip photonic synapse Zengguang Cheng CR, 1 Wolfram H. P. Pernice,2 C. David Wright,3 Harish Bhaskaran1*. On-chip photonic synapse. 2017.
28. Yu J, Yang X, Gao G, Xiong Y, Wang Y, Han J*, et al.* Bioinspired mechano-photonic artificial synapse based on graphene/MoS<sub>2</sub> heterostructure. *Science Advances* 2021, **7**(12): eabd9117.
29. Hazan A, Ratzker B, Zhang D, Katiyi A, Sokol M, Gogotsi Y*, et al.* MXene-Nanoflakes-Enabled All-Optical Nonlinear Activation Function for On-Chip Photonic Deep Neural Networks. *Adv Mater* 2023, **35**(11): e2210216.
30. Zhang P, Xiao X, Ma ZW. A review of the composite phase change materials: Fabrication, characterization, mathematical modeling and application to performance enhancement. *Applied Energy* 2016, **165:** 472-510.
31. Zhang L, Dong J, Ding F. Strategies, Status, and Challenges in Wafer Scale Single Crystalline Two-Dimensional Materials Synthesis. *Chemical Reviews* 2021, **121**(11): 6321-6372.
32. Ma H, Xiao X, Wang Y, Sun Y, Wang B, Gao X*, et al.* Wafer-scale freestanding vanadium dioxide film. *Science Advances* 2021, **7**(50): eabk3438.
33. Shi Y, Ren J, Chen G, Liu W, Jin C, Guo X*, et al.* Nonlinear germanium-silicon photodiode for activation and monitoring in photonic neuromorphic networks. *Nat Commun* 2022, **13**(1): 6048.
34. Li H, Wu B, Tong W, Dong J, Zhang X. All-Optical Nonlinear Activation Function Based on Germanium Silicon Hybrid Asymmetric Coupler. *IEEE Journal of Selected Topics in Quantum Electronics* 2022: 1-1.
35. Wu B, Li H, Tong W, Dong J, Zhang X. Low-threshold all-optical nonlinear activation function based on a Ge/Si hybrid structure in a microring resonator. *Optical Materials Express* 2022, **12**(3): 970.
36. Jha A, Huang C, Prucnal PR. Reconfigurable all-optical nonlinear activation functions for neuromorphic photonics. *Opt Lett* 2020, **45**(17): 4819-4822.
37. Wang B, Yu W, Duan J, Yang S, Zhao Z, Zheng S*, et al.* Microdisk modulator-assisted optical nonlinear activation functions for photonic neural networks. *Optics Communications* 2023: 130121.
38. Ghosh J, Nag A. An Overview of Radial Basis Function Networks. In: Howlett RJ, Jain LC (eds). *Radial Basis Function Networks 2: New Advances in Design*. Physica-Verlag HD: Heidelberg, 2001, pp 1-36.
39. Srivastava N, Hinton G, Krizhevsky A, Sutskever I, Salakhutdinov R. Dropout: a simple way to prevent neural networks from overfitting. *The journal of machine learning research* 2014, **15**(1): 1929-1958.
40. LeCun Y, Touresky D, Hinton G, Sejnowski T. A theoretical framework for back-propagation. Proceedings of the 1988 connectionist models summer school; 1988; 1988. p. 21-28.
41. LeCun Y, Bottou L, Bengio Y, Haffner P. Gradient-based learning applied to document recognition. *Proceedings of the IEEE* 1998, **86**(11): 2278-2324.
42. Li GHY, Sekine R, Nehra R, Gray RM, Ledezma L, Guo Q*, et al.* All-optical ultrafast ReLU function for energy-efficient nanophotonic deep learning. *Nanophotonics* 2022,

**0**(0).

43. Liu B, Xu S, Ma B, Yi S, Zou W. Low-threshold all-optical nonlinear activation function based on injection locking in distributed feedback laser diodes. *Optics Letters* 2023, **48**(15)**:** 3889.

44. Tavanaei A, Ghodrati M, Kheradpisheh SR, Masquelier T, Maida A. Deep learning in spiking neural networks. *Neural networks* 2019, **111:** 47-63.

45. Miller DAB. Attojoule Optoelectronics for Low-Energy Information Processing and Communications. *Journal of Lightwave Technology* 2017, **35**(3)**:** 346-396.

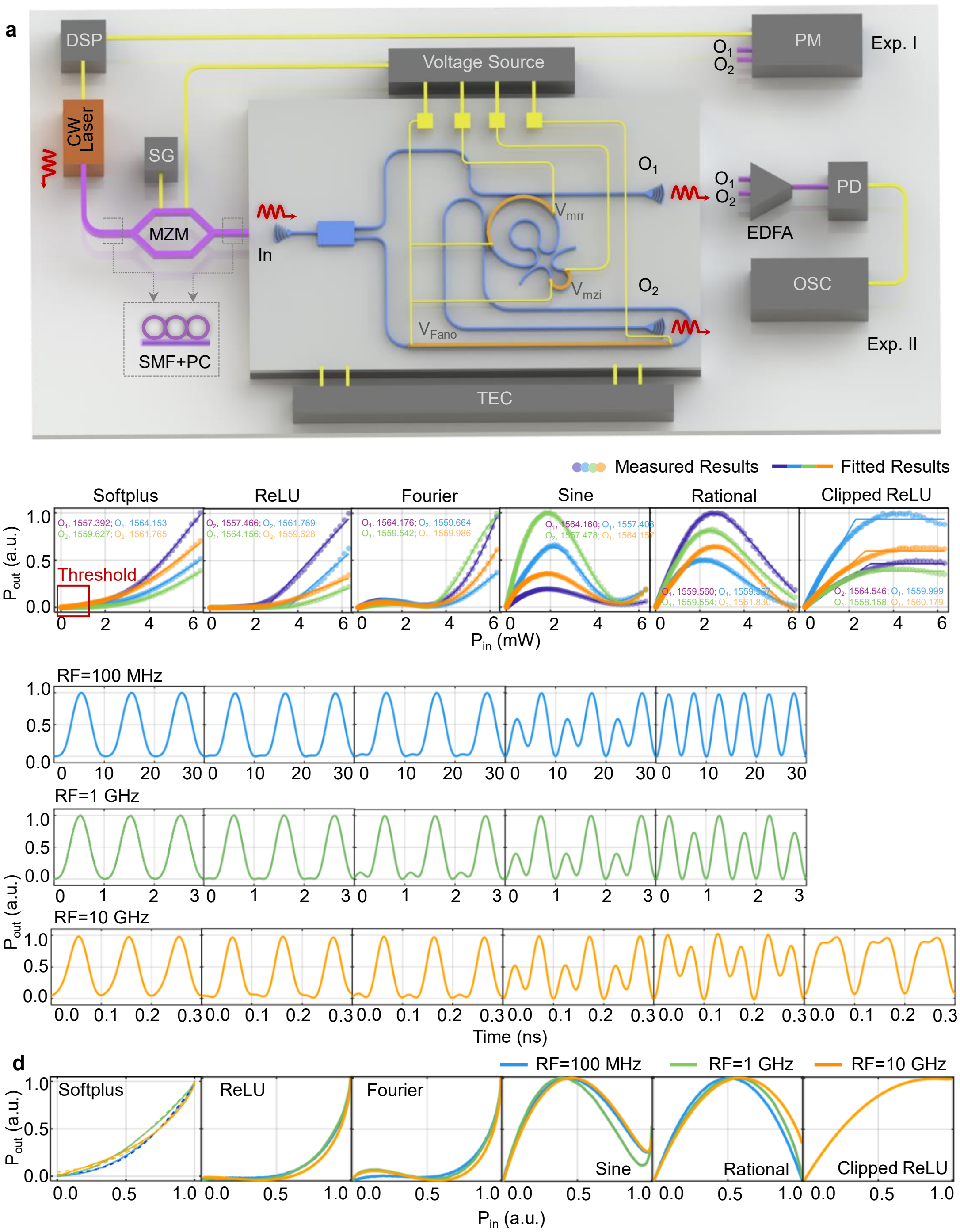

**Fig. 1. Characterization of the Fano-based optical nonlinear activator. a,** Experimental setup for activation function generation under static characterization (Exp. I) and high-speed responses (Exp. II). DSP, digital signal processing; PM, power meter; CW, continuous-wave; SG, signal generator; MZM, Mach-Zehnder modulator; SMF, single-mode fiber; PC, polarization controller; TEC, thermo-electric cooler; EDFA, erbium-doped fibre amplifier; PD, photodetector; OSC, oscilloscope. **b,** Measured and fitted $P_{out}$- $P_{in}$ relations under static characterization. **c,**

Output time-domain waveforms monitored on the sampling OSC of modulation frequencies (100 MHz~10 GHz). **d,** Measured and fitted $P_{out}$- $P_{in}$ relations under high-speed responses.

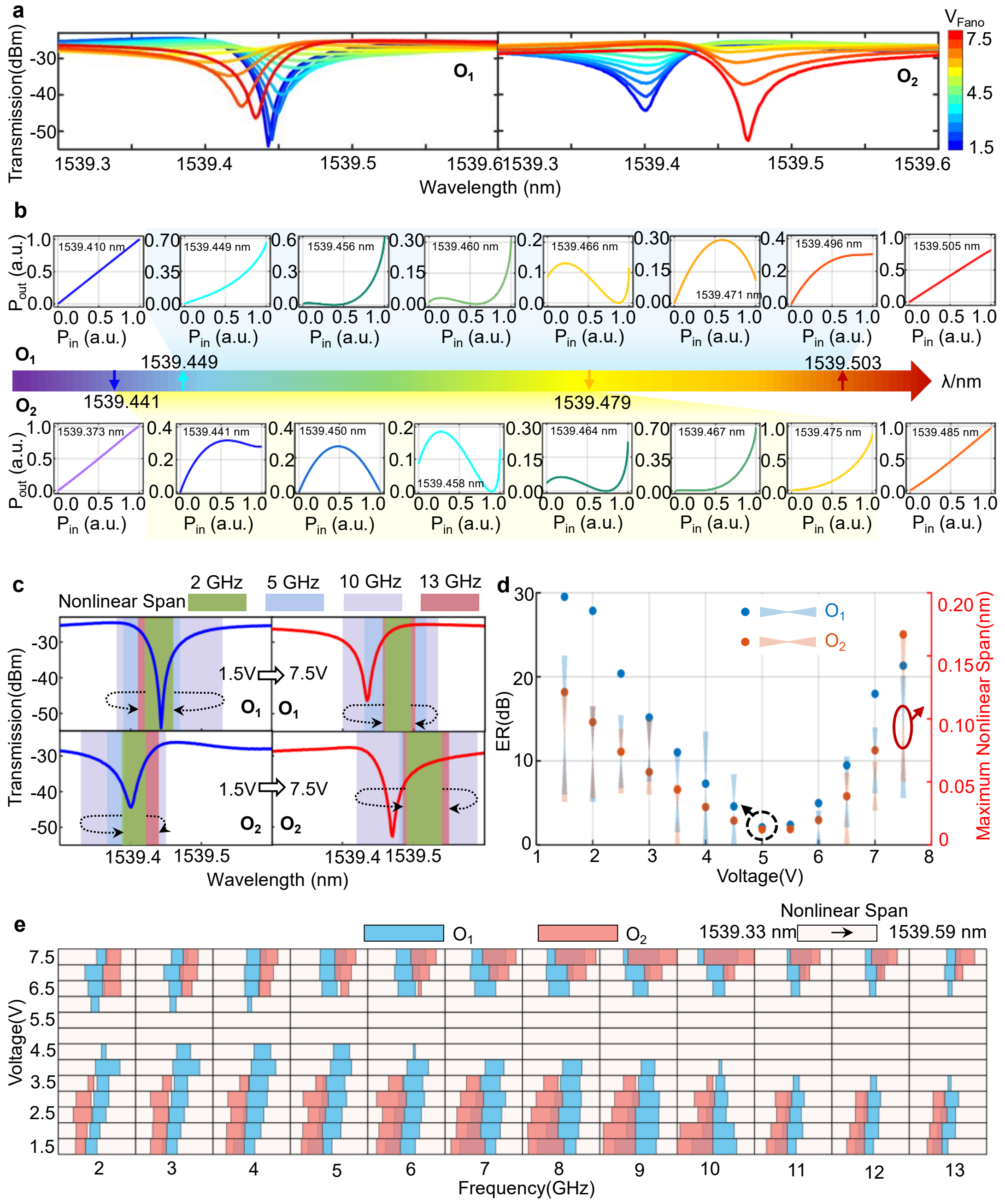

**Fig. 2. Revolution progress for optical nonlinear activation functions. a,** Transmission spectra at the device's two output ports with Fano biases from 1.5 V to 7.5 V. **b,** Instantaneous $P_{out}$-$P_{in}$ relations at two output ports under Fano bias fixed at 1.5 V and the modulation frequency set as 10 GHz. **c,** The tuning process of nonlinear spans from 2 GHz to 13 GHz at the two-end Fano biases. **d,** Extinction ratio and maximum nonlinear span at different voltages

of Fano bias. **e,** A full look-up table of nonlinear span at two output ports for a two-dimensional sweep of the modulation frequency and Fano bias.

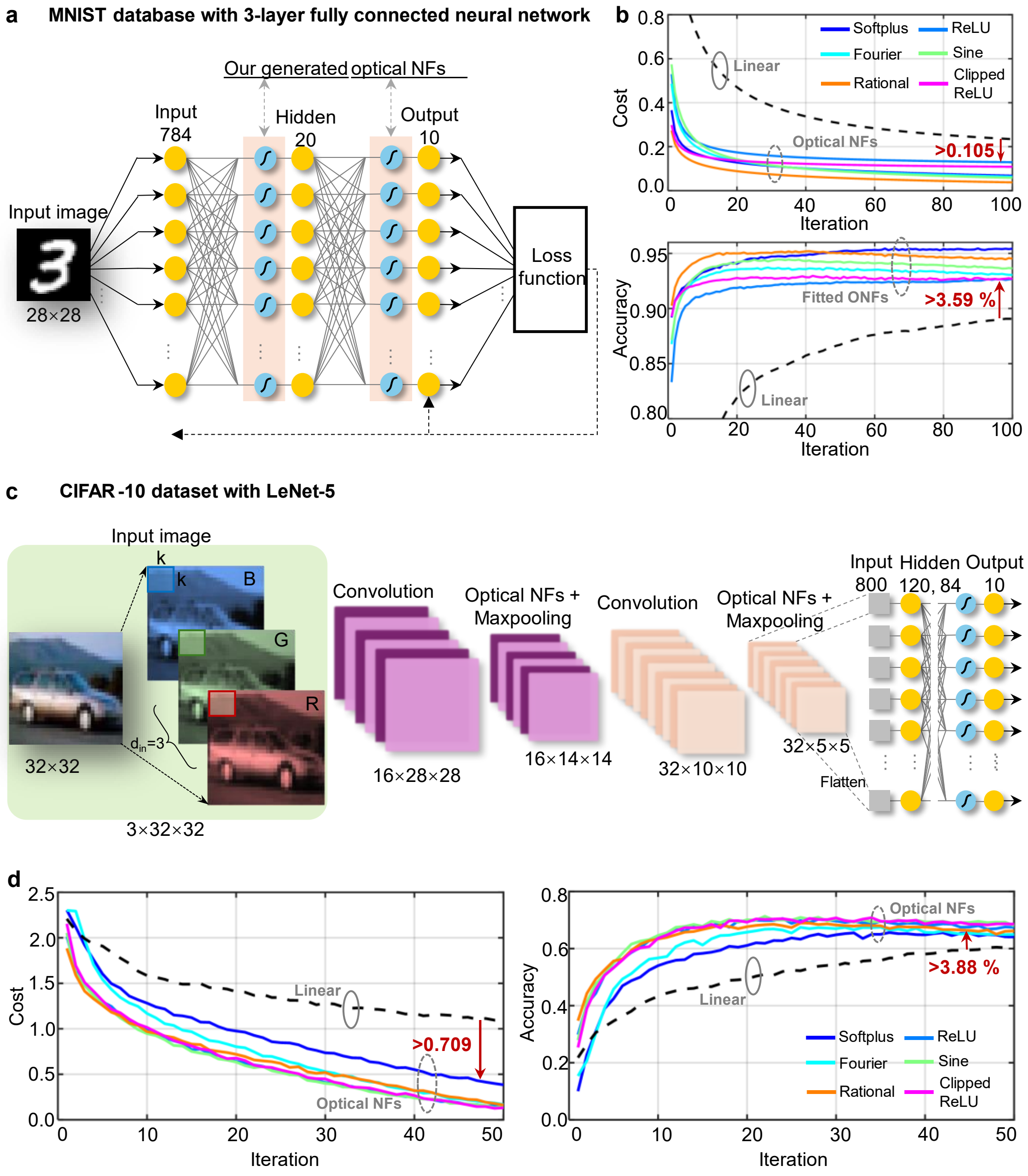

**Fig. 3 PNN training and results. a**, MNIST dataset with 3-layer fully connected neural network **b,** Convergence cost and accuracy using different fitted optical NFs. The comparison baseline is trained with linear function. **c**, CIFAR-10 dataset with LeNet-5. **d,** Convergence cost and accuracy using six classes of fitted optical NFs.

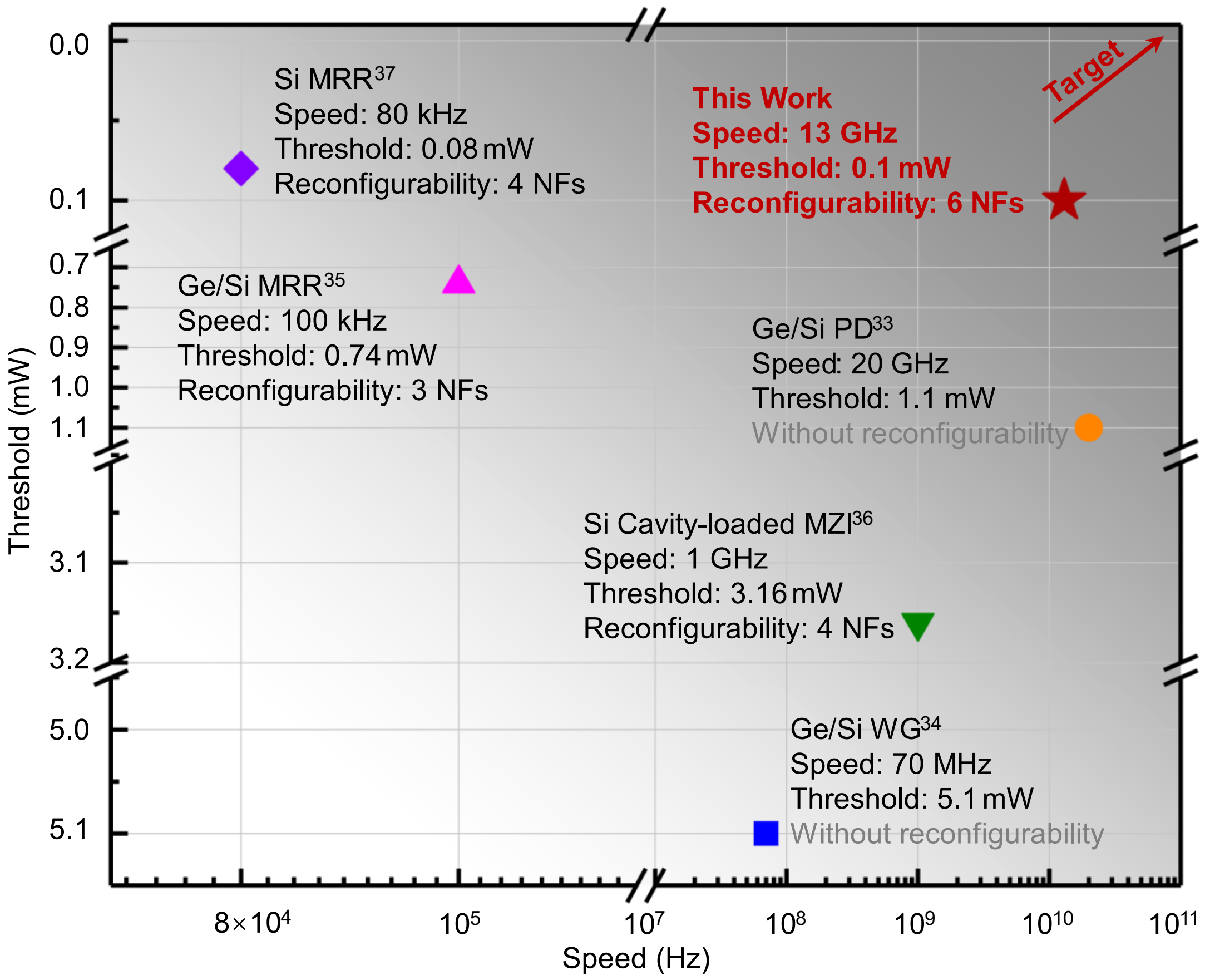

**Fig. 4. Performance comparison of different silicon-based AONA structures with continuous wave inputs.**